\documentclass[epj]{svjour}
\usepackage{makeidx}
\usepackage{graphicx}
\usepackage{amssymb}
\usepackage{lscape}
\usepackage{cite}

\begin{document}

\authorrunning{Rusev, Grosse et al.}

\titlerunning{Pygmy dipole strength in Mo-isotopes}

\title{Pygmy dipole strength close to particle-separation energies - \\the case 
of the Mo isotopes}

\author{Gencho Rusev\inst{1}, Eckart Grosse\inst{1, 2}, Martin Erhard$^1$, Arnd 
Junghans$^1$, Krasimir Kosev$^1$, Klaus-Dieter Schilling$^1$, Ronald 
Schwengner$^1$ \and Andreas Wagner$^1$}

\institute{Forschungszentrum Rossendorf, 
Institut f\"ur Kern- und Hadronenphysik,\\
Postfach 510119, 01314 Dresden Germany \and Technische Universit\"at Dresden, 
Institut f\"ur Kern- und Teilchenphysik,\\
01062 Dresden, Germany}

\date{Received: date / Revised version: date}

\abstract{The distribution of electromagnetic dipole strength in 
$^{92,\,98,\,100}$Mo has been investigated by photon scattering using 
bremsstrahlung from the new ELBE facility. The experimental data for well 
separated nuclear resonances indicate a transition from a regular to a chaotic 
behaviour above 4\,MeV of excitation energy. As the strength distributions 
follow a Porter-Thomas distribution much of the dipole strength is found in 
weak and in unresolved resonances appearing as fluctuating cross section. An 
analysis of this quasi-continuum - here applied to nuclear resonance 
fluorescence in a novel way - delivers dipole strength functions, which are 
combining smoothly to those obtained from ($\gamma$,\,n)-data. Enhancements at 
6.5 MeV and at $\sim 9\,$MeV are linked to the pygmy dipole resonances 
postulated to occur in heavy nuclei.}


\maketitle

\section{Dipole strength in heavy nuclei}
\label{sec:7.1}
The response of nuclei to dipole radiation is of special importance for the 
synthesis of the chemical elements in the cosmos: Particle thresholds may be 
crossed in hot or explosive scenarios leading to the production of new nuclides 
from previously formed heavier ones by dissociation in the thermal photon bath. 
This is likely to be the main path for the generation of the approximately 
30\,-\,40 neutron-deficient nuclides which cannot be produced in neutron capture 
reactions \cite{1}. For the understanding and modelling of this so-called 
p-process the dipole strength function up to and near the particle thresholds 
has to be known accurately \cite{2}. As shown previously \cite{3}, details of 
the dipole strength (now in n-rich nuclei) may as well have large consequences 
for the r-process path and also s-process branchings are influenced by nuclear 
excitations \cite{4} induced by thermal photons.\\
The experimental knowledge \cite{5} on dipole strength is reasonably well 
established for many heavy and medium mass nuclei in the region of the giant 
dipole resonance (GDR) by ($\gamma$,\,xn) studies, which often also cover the 
region directly above the neutron threshold $S_n$. At lower energies three 
features have been discussed to be of importance for processes in 
high-temperature cosmic environments:\\
a) the fall-off \cite{6, 7, 8, 9} of the E1-strength on the low-energy slope of 
the GDR;\\ 
b) the E1-strength between the ground-state (gs) and low energy excitations and 
its proper extension \cite{10, 11, 12} into the regime (a);\\
c) the occurrence of additional pygmy-resonances, \cite{3, 13, 14, 15}, which 
are assumed to be not as broad as the GDR, but wider as compared to the average 
level distance $D$ - thus forming an intermediate structure enclosing many 
levels. Their low energy may well enhance their contribution to 
photo-dissociation processes in spite of their relatively low strength as 
compared to the GDR.\\
In principle, also M1-transitions contribute to the dipole strength, but the 
average M1 strength is typically 1-2 orders of magnitude smaller as compared to 
E1, and they are frequently \cite{1, 2, 3} not taken into account.\\
A further approximation has to be introduced to estimate dipole strengths for 
transitions not connected to the ground state; there are two possibilities 
proposed in the literature:\\
a) Strictly following a hypothesis set up by Brink and Axel \cite{16} 
the strength of a transition connecting in a given nucleus two levels separated 
by an energy difference $E_\gamma$ only depends on $E_\gamma$, on the transition 
type and on a statistical factor determined by the two spins.\\
b) As proposed by Kadmenskii \cite{10}, a temperature dependence of the strength 
function is introduced, which smears out the GDR-strength into the region below 
and effectively connects a certain excitation region above the ground state to 
the GDR domain.\\
A combination of the two prescriptions has been tried \cite{12}, but this 
suffers from an inconsistency which may be of principal nature: At low energy 
the nuclear excitation is quantized, whereas with increasing energy statistical 
concepts from thermodynamics are more effective in describing the increasing 
complexity. Microscopic calculations \cite{8, 11} may allow to develop a 
consistent transition from the low to the high excitation region near the GDR 
and to properly account for other intermediate strength. 
 
\section{Nuclear resonance fluorescence} 
\begin{figure}
\resizebox{0.4\textwidth}{!}{\includegraphics{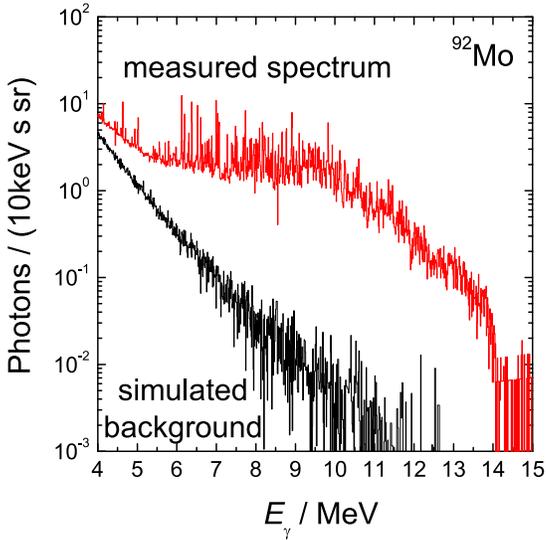}}
\caption{Spectrum of bremsstrahlung photons scattered by $^{92}$Mo into 
$127^{\circ}$. Above a background caused by atomic processes - whose height was 
determined from a Monte-Carlo simulation - an accumulation of sharp lines near 
7\,MeV is observed as well as a strong quasi-continuum extending up to the 
endpoint energy of $\sim$ 14\,MeV. Note that lines from identified background 
sources have been subtracted.}
\label{fig:1}
\end{figure}

\begin{figure}
\resizebox{0.4\textwidth}{!}{\includegraphics{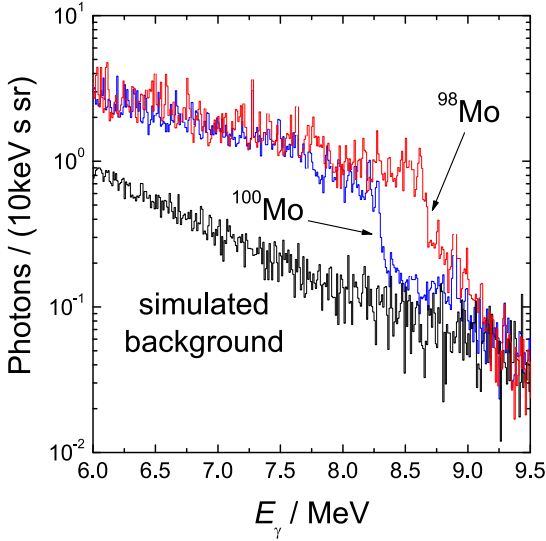}}
\caption{Same as Fig.\,1 for $^{98}$Mo and $^{100}$Mo. The background 
simulation is the same for both isotopes. This allows a test of the 
determination of the continuum in the region between the two (different) 
neutron thresholds.}
\label{fig:2}
\end{figure}

Information about energy dependent dipole strength functions can be obtained 
from data on multi-step gamma-decays following n-capture \cite{18} or direct 
reactions \cite{12}, from inelastic electron scattering \cite{19} or elastic 
photon scattering \cite{14, 15, 20}. For the region above the n-threshold the 
electromagnetic strength in a very large number of stable nuclei has been 
experimentally determined by observing the neutrons emitted after the excitation 
by quasi-monochromatic photons\cite{5, 21}. Data on this $(\gamma$,\,xn) 
process, taken as averages over a certain energy bin, allow to determine an 
(averaged) dipole strength function $f_1$ in this region by making use of the 
relation:\\
\begin{equation}
f_{1}(E) = (3\pi^{2}\hbar^{2}c^{2}E)^{-1}\cdot \sigma_{\gamma}(E) 
\label{eq:1}
\end{equation}
where $\sigma_{\gamma}$ describes the dipole dissociation of a spin 0 target by 
a photon of energy $E$. The strength function data obtained from the other 
methods suitable for the lower excitation energy should connect smoothly to 
$(\gamma$,\,xn) data to yield information about the dipole strength over a wide 
energy range from the ground state up to far above the particle emission 
thresholds. Of special importance here is the much discussed 
\cite{7, 10, 11, 12} question, if the extrapolation of the Lorentzian fit to the 
GDR fits dipole strength data also at and below the neutron threshold.\\  
One experimental method has delivered interesting information about nuclear 
dipole strength; results from it have become more and more detailed with the 
improving measurements. Using a bremsstrahlung beam the scattering of photons 
with energies up to the GDR is observed by large volume Ge-detectors. The good 
resolution (3\,-\,5\,keV) of such detectors in combination with Compton 
suppression shields limit the detector response matrix such that it becomes 
nearly completely diagonal. Thus the signal from elastically scattered photons 
identifies the energy of the incoming photon out of the bremsstrahlung 
continuum, as all nuclear levels with sufficient transition strength to the 
ground-state are observed as narrow elastic scattering resonances up to the 
respective neutron emission thresholds; inelastic scattering to levels above the 
target ground states is also observed \cite{22} and has to be considered in the 
data analysis.\\
From the knowledge of inelastic scattering via higher lying levels the feeding 
to a certain level can be accounted for in the analysis of the elastic cross 
section $\sigma_{\gamma\gamma}$:
\begin{equation}
I(E_R) = \int_R\sigma_{\gamma\gamma}(E)dE-I_{feed}
\label{eq:2}
\end{equation}
with the integral taken over the (narrow) resonance R centered at 
$E_{\gamma}=E_R$. The ground state (spin 0) width $\Gamma_0$ of such a resonance 
(spin 1) and $I$ are related by 
\begin{equation}
I(E_R) = 3\Big(\frac{\pi \hbar c}{E_R}\Big)^2\cdot \frac{\Gamma_{0}^{2}}{\Gamma}
\label{eq:3}
\end{equation}
where $\Gamma=\Gamma_{0}+\Gamma_{c}$ is its total width and $\Gamma_{c}$ the 
summed width of all decay processes competing to the decay back into the ground 
state. The contribution of the $N_{\Delta}$ resonances in an energy interval 
$\Delta$ to the dipole strength function can be calculated by:
\begin{equation}
f_{1, \Delta} = \frac{1}{\Delta}\sum_{i=1}^{N_\Delta}\frac{\Gamma_{0, i}}
{E_{R, i}^{3}}
\label{eq:4}
\end{equation}
The present study on the Mo isotopes uses eq. (\ref{eq:4}) only for the single 
line spectra \cite{22} observed below $E_R\simeq 4$\,MeV with low endpoint 
bremsstrahlung, the other data require a more sophisticated analysis.\\
A non negligible contribution to the scattering results from non-nuclear 
processes as Compton scattering with the subsequent bremsstrahlung reaching the 
detector; pair production leads to photon background in a similar way. Such 
processes can be simulated \cite{20} to high accuracy and subtracted from the 
data as shown in Figs. \ref{fig:1} and \ref{fig:2}, to yield the "true" nuclear 
resonance fluorescence (nrf) cross section $\sigma_{\gamma\gamma}$ to be used in 
eq. (\ref{eq:2}). Only the resulting difference spectra contain information 
about the nuclear dipole strength. As the direct contributions from Thomson and 
Delbr\"uck scattering are weaker by orders of magnitude, the scattering of MeV 
photons is primarily nrf. Only when $\Gamma_c$ completely dominates $\Gamma$ - 
e.g. above the neutron threshold $S_n$ - the contribution of these direct 
scattering processes can no longer be neglected. Contributions to the spectra 
from higher multipole radiation are identified by a differing angular 
distribution. M1-transitions can be identified through the use of linearly 
polarized photons \cite{24, 25, 26, 27}. 

\section{Photon scattering experiments on Mo-isotopes}
The present paper reports on photon scattering experiments for the Mo isotopes 
with $A$\,=\,92,\,98 and 100. This rather wide range in neutron number $N$ may 
allow a reasonable extrapolation to unstable isotopes; for the case of the pygmy 
resonance the E1-strength has been predicted \cite{13} to vary strongly with $N$ 
whereas its energy should weakly depend on $N$. The low-$N$ stable Mo isotopes 
are p-process nuclides with a surprisingly high cosmic abundance thus making 
accurate information on the response of Mo-isotopes to dipole radiation 
especially desirable. The photon scattering experiment at the new radiation 
source ELBE with its superconducting electron linac \cite{23} was set up 
similar to previous nrf-studies \cite{13, 14, 15, 28}. One of its special 
features is, that the bremsstrahlung emerges from a thin Nb-foil (approx. 
5 mg/cm$^2$) bombarded with a beam of approx. $650\,\mu{\rm A}\,=\,4\cdot 
10^{15}{\rm e/s}$. The electron beam is deflected into a well shielded beam dump 
after passing the radiator. Highly enriched targets of $^{92}$Mo, $^{98}$Mo and 
$^{100}$Mo have been used with masses of 2-3\,g each. Electrons with an average 
momentum of 14\,MeV/c and an rms momentum spread of 0.07\,MeV/c were used in 
these experiments. For each target a second run with a lower electron momentum 
was performed such that the endpoint of the bremsstrahlung continuum stayed 
below the neutron and proton emission thresholds. The photon beam was limited 
transversely by a 2.5\,m long conical Al-collimator, such that an approximate 
photon flux of about $10^{7}\gamma/({\rm s}\cdot {\rm cm}^{2}\cdot$ MeV) hits 
the experimental targets. Four large high-purity Germanium semiconductor 
detectors (enclosed by anti-Compton shields from BGO) set up at 90$^{\circ}$ and 
127$^{\circ}$ registered the photons scattered by the target.\\
The detectors were shielded from unwanted background radiation by lead bricks 
positioned around the photon beam dump, near the collimator exit and around 
the beam tube at the target. A conical opening between target and Ge-detector, 
which was filled by only 2 cm of Pb to absorb the very intense low energy 
photons, determined the angle of observation and the solid angle. Details of the 
set-up and of the Monte-Carlo simulations performed with the aim to optimize it 
are described elsewhere \cite{20}. According to these simulations the main 
background contribution to the spectra of scattered photons (as shown in 
Fig. \ref{fig:1}) is due to bremsstrahlung produced from pair production and 
Compton scattering in the target. The simulation of this background (cf. Fig. 
\ref{fig:1} and \ref{fig:2}) could be made sufficiently accurate to allow the 
generation of pure nrf-spectra by subtracting the simulated non-resonant 
contribution from the experimental data.\\
Due to the 300\,keV difference of the neutron binding energies of $^{98}$Mo and 
$^{100}$Mo a subtraction of the $^{100}$Mo data from the $^{98}$Mo data results 
in a pure nrf-spectrum in the range $S_{n}(^{100}$Mo) to $S_{n}(^{98}$Mo) - 
under the suggestive assumption that the non resonant background is the same for 
both isotopes (see Fig. \ref{fig:2}). From the fact, that this procedure leads 
within errors to the same nrf-strength in this energy bin as the Monte-Carlo 
based subtraction, a test of the accuracy of the latter procedure is obtained. 
It should be noted here, that the subtraction explained above was performed with 
the data after their complete correction for detector response. Thanks to the 
high full energy efficiency of the Ge-detectors used and as result of the good 
Compton suppression by the BGO shields such a response correction could be 
performed without introducing large statistical uncertainties.
\begin{figure}
\resizebox{0.3\textwidth}{!}{\includegraphics{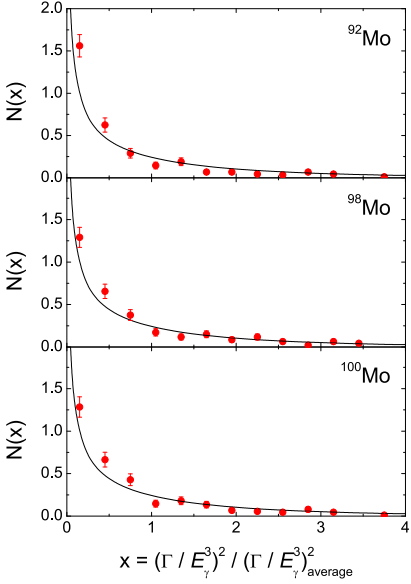}}
\caption{Distributions of the transition widths as determined from each 50 
transitions starting at 4\,MeV, all reduced by the phase space factor 
$E_\gamma^{3}$. The drawn lines depict Porter-Thomas distributions.}
\label{fig:3}
\end{figure} 

\section{Dipole strength in isolated narrow resonances}
\begin{figure}
\resizebox{0.3\textwidth}{!}{\includegraphics{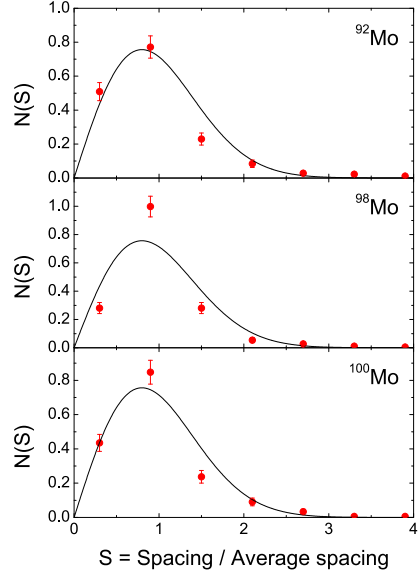}}
\caption{Next neighbour distance distributions of the isolated narrow resonances 
above 4\,MeV. Averages were formed over each 50 transition energies and the 
ratios of distance to average distance were collected in 7\,bins. The drawn 
lines represent a Wigner distribution.}
\label{fig:4}
\end{figure}

In the nrf-spectra obtained from the raw data as shown in Figs. \ref{fig:1} and 
\ref{fig:2} many isolated resonances could be analysed in the range from 4\,MeV 
up to the endpoint energy of 13.2\,MeV; the lower part of the spectra is 
discussed elsewhere \cite{22}. For the three isotopes 299, 310 and 296 
resonances, respectively, could be distinguished above 4\,MeV; no strong lines 
can be identified above the respective neutron threshold. The ratios of the 
intensities 
\begin{equation}
\frac{dI}{d\Omega}=\int_R\frac{d\sigma(E)}{d\Omega}dE,
\label{eq:5}
\end{equation}
where the integral includes the full (narrow) resonance R, as observed at 
scattering angles 90$^{\circ}$ and 127$^{\circ}$ have been compared to the 
values expected for a spin sequence 0-1-0 or 0-2-0 for excitation and 
deexcitation. Apparently nearly all of the transitions are due to spin 1 
resonances. They are assumed to be E1, as from a previous experimental nrf-study 
on $^{92}$Mo only 1 resonance is identified \cite{24} to have positive parity; 
its M1-strength to the gs corresponds to $0.23\mu_{N}^{2}$: As in the 
neighbouring nucleus $^{90}$Zr a total M1 strength of $6.7\mu_{N}^{2}$ was found 
\cite{25} to lie between 8 and 11\,MeV, it is likely that a few more M1 
resonances are also present in $^{92}$Mo, but in any case most of the 
nrf-strength clearly is of E1 character. A similarly low M1 strength has been 
observed for $^{116}$Sn and $^{124}$Sn \cite{27} and proposed for $N$\,=\,82 
nuclei \cite{15}.\\
For the well identified resonances the reduced width distributions are displayed 
for the three isotopes in Fig. \ref{fig:3} after a normalization to the 
respective average width taken over bins containing 50 dipole resonances each, 
observed in the range from 4 MeV up to the neutron binding energies. The 
distributions are in full agreement to Porter-Thomas distributions indicating 
chaotic statistics in the ground state transition strengths.\\
To study as well the statistical properties of the nearest neighbour spacings we 
also treat the resonances above 4~MeV in groups of 50 and determine the average 
spacing in each group. The actual spacings divided by this average are shown as 
black dots in Fig. \ref{fig:4} in comparison to Wigner distributions.\\
From comparison to the data taken at lower end-point energy it is obvious that 
above 5\,MeV nearly all of the identified transitions connect to the ground 
state. To obtain an estimate of the possible corrections necessary to account 
for the incorrect interpretation of non-gs transition energies as level energies 
we have performed respective Monte Carlo simulations of level sequences 
describing a Porter-Thomas or a Poisson case. Only a small distortion is caused, 
when the transition energies resulting from these simulations are (eventually 
erroneously) treated as level energies. In any case, our Mo data do closely 
resemble Wigner distributions, again pointing to chaotic statistics.\\

\section{Level densities and fluctuating cross sections}
The high level density in combination with Porter-Thomas fluctuations cause a 
large portion of the strength to appear in many weak transitions which are 
likely to be missed experimentally. This is why an average level density in its 
dependence on the excitation energy can only be determined from a fluctuation 
analysis on the basis of Porter-Thomas statistics \cite{19}. Similarly, the 
dipole strength in a certain energy interval has to be obtained by integrating 
the complete nrf spectra - i.e. all counts in discrete resonances and in the 
quasi-continuum in between, after subtraction of the non-resonant background. 
The accuracy of the determination of this background can be judged from what is 
presented in Fig. \ref{fig:2} to be sufficiently high. As obvious from the 
analysis of the data shown in Figs. \ref{fig:1} and \ref{fig:2}, the average 
strength in the last MeV below $S_n$ is approximately increased by at least a 
factor 3 when the continuum is included.\\
Another important correction has to account for inelastic scattering, i.e. 
transitions branching to excited states. Its effect can be identified from data 
taken at different endpoint energies of the bremsstrahlung spectrum. When the 
$^{100}$Mo data taken with an endpoint energy of 8.3\,MeV are compared to the 
data reaching up to the threshold of 7.8\,MeV, the intensity distribution 
originating from these extra 500\,keV of bremsstrahlung can be identified: More 
than 50\,\% of this intensity is observed as cascades with photons in the range 
of 3 of 5\,MeV. Obviously the remaining intensity observed as gs-transitions has 
to be multiplied by a factor of 2-3 to obtain the full excitation strength. To 
obtain an approximation for this correction factor the assumption \cite{28, 29} 
is adopted, that below $S_n$ and $S_p$ inelastic processes (i.e. branching) can 
on the average be accounted for by setting $\Gamma_c$\,=\,0.2\,eV (for the Zr 
region) in combination with the level density \cite{30}. A more accurate 
correction is to be gained from the experiments at lower energy and by 
Monte-Carlo-simulations; obviously the approximation applied as described above 
cannot introduce extra structures. \\
This is the first time, that high resolution nrf-spectra are analyzed such that 
not only the isolated resonances are included, but also the fluctuating 
quasi-continuum. This is accomplished by calculating the dipole strength 
function for the energy range up to $S_n$ (as covered in this experiment) from 
the elastic component $\sigma_{\gamma \gamma}$ of nrf:
~\\[-2cm]
\begin{equation}
f_{1}(E) = 
(3\pi^{2}\hbar^{2}c^{2}E)^{-1}\cdot\frac{\Gamma}{\Gamma_0}\cdot
\frac{1}{\Delta}\cdot
\int_{\Delta}\sigma_{\gamma\gamma}(E_{\gamma})\cdot dE_\gamma
\label{eq:6}
\end{equation} 
~\\[-2cm]
where $\Delta$ is the interval selected around $E$ for averaging the widths 
$\Gamma_0$ and $\Gamma$ and the photon energy $E_\gamma$. The photon absorption 
cross section is thus derived from the observed elastic photon scattering cross 
section $\sigma_{\gamma \gamma}$ by correcting bin-wise for inelastic 
scattering (i.e. branching). This determination of $f_1$ then allows a 
quantitative comparison of nrf - and ($\gamma$,\,xn) - data (cf eq. \ref{eq:1}), 
and both can be directly combined to extract a continuous dipole strength 
function.

\begin{figure}
\resizebox{0.7\textwidth}{!}{\includegraphics{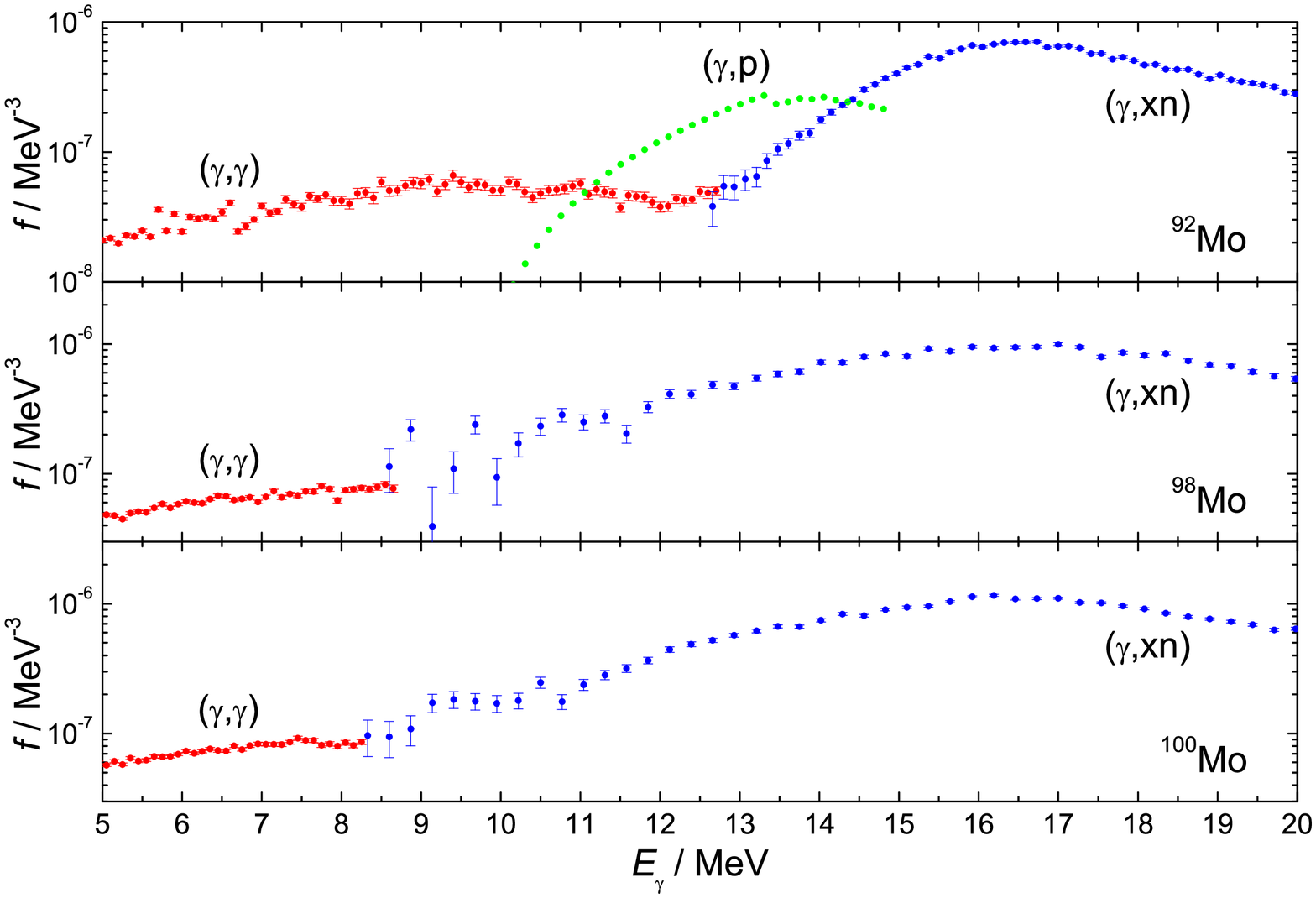}}
\end{figure}
\begin{minipage}[]{0.5\textwidth}
\parbox{0.4\textwidth}{\hfill}
\parbox{0.5\textwidth}{~\\{\bf Fig.\,5.} Dipole strength functions determined 
from the photon scattering and the ($\gamma$,\,xn)-data as described in the 
text. The nrf data are shown in bins of 100 keV; this makes the enhancement of 
strength in several single resonances near 6\,MeV less obvious. In the case of 
$^{92}$Mo also the ($\gamma$,\,p)-process has to be accounted for; this is 
indicated by including a respective cross section calculation \cite{31} for the 
corresponding contribution to be added to obtain the total strength. Due to the 
weakness of quadrupole excitations the plotted $f$ is effectively the dipole 
strength function $f_1$.}
\vspace*{5cm}
\end{minipage}
~\\[-5.5cm]~\\[1cm]

\section{The distribution of photon strength and pygmy resonances}
The good compatibility of the nrf-strength (corrected for branching) directly 
below $S_n$ and the ($\gamma$, xn)-data directly above encourages a search for 
structure in $f_1(E)$ derived from the two data sets. The data indicate an 
enhancement of the dipole strength at $\sim$9\,MeV. In $^{92}$Mo this possible 
pygmy resonance appears below and in $^{100}$Mo it is above $S_n$; in $^{98}$Mo 
the region directly above $S_n$\,=\,8.6\,MeV shows some irregularities. In an 
old tagged photon scattering experiment on natural Zr cross section enhancements 
at 9.1 and 11.6\,MeV were found \cite{28}; from the isotope enrichment and the 
n-threshold values it is argued, that they should originate from $^{90}$Zr. Most 
of that strength was shown not to be M1 \cite{25, 29}. The Mo-data from ELBE as 
well as these results have to be compared to broad resonance-like structures 
seen \cite{27} in Sn isotopes at 6.7 and between 8.0 and 8.7\,MeV. The strength 
as extracted from an experiment \cite{27} on $^{116}$Sn an $^{124}$Sn clearly 
stays below the extrapolation of the GDR-Lorentzian as only narrow isolated 
resonances had been analyzed. Although the broad structures seen in Sn by the 
tagged photon study \cite{28} and in the new Ge-detector experiment \cite{27} 
appear at nearly the same energies, the strength observed in the region of the 
broad pygmy structure differs by a factor of  two between the two types of 
experiment. Obviously, an analysis of the well isolated peak on the basis of eq. 
(\ref{eq:4}) misses much of the strength which is accounted for by following a 
procedure characterized by eq. (\ref{eq:6}), which does not ignore the 
fluctuating quasi-background. The only other isotope chain studied in this range 
of $A$, the even Ge-isotopes, show \cite{26} no clear pygmy resonance appearing 
in the discrete spectra below $S_n$. The average strength function obtained from 
these resonances by using formula (\ref{eq:4}) amounts to 
$\sim 10^{-8}$MeV$^{-3}$ and clearly stays below the extrapolated Lorentzian 
extracted from the ($\gamma$,\,xn)-data.\\
Recent HFB-QRPA calculations \cite{8} give good fits to GDR data when the force 
Bsk-7 of Skyrme-type is used to describe the effective nucleon-nucleon 
interaction. Nevertheless it should be noted here, that $\sigma(\gamma$,\,n) 
directly above threshold is well described only for $^{94}$Mo, whereas the 
calculation is below the experimental value \cite{21} for $^{100}$Mo by a 
factor 4. An extension of such calculations to lower $E$ - including the pygmy 
region - seems interesting.
\section{Conclusion}
The response of nuclei to dipole radiation can well be studied by photon 
scattering investigated at a bremsstrahlung facility like ELBE. Using a 
sufficiently high endpoint energy and correcting the data for inelastic 
processes allows to directly combine the dipole strength functions $f_1$ 
obtained from the ($\gamma, \gamma$) (i.e. nrf) and the ($\gamma$,\,xn) data; 
together they span the full range from the ground state to the GDR. A comparison 
of $f_1(E)$ to a Lorentzian extra\-polated from the GDR needs a more thorough 
discussion of the spreading of the GDR than is accomplished by just fitting near 
its maximum \cite{5, 21}. Apparently the large apparent width of the GDR in 
$^{100}$Mo may be caused by a deformation of that nucleus; accounting for that 
by a two-resonance-fit would reduce the low energy tail considerably. 
Calculations of the type presented recently \cite{2, 8, 11} may help to clarify 
this point, especially when the nuclear deformation is included with sufficient 
accuracy.\\
Above $\sim$4\,MeV the predicted level densities \cite{30} increasingly surpass 
the number of identifiable resonances and apparently the levels of $^{92}$Mo, 
$^{98}$Mo and $^{100}$Mo show signs of a chaotic structure: The next neighbour 
distance distributions of the clearly identified peaks are Wigner distributed 
and their ground-state transition widths follow Porter-Thomas distributions. 
From these facts one expects the photon scattering excitation functions - which 
in a bremsstrahlung beam are observed simultaneously over a wide range - to show 
Porter-Thomas fluctuations in case the detector resolution surpasses the average 
peak distance. As this is the case for most of the Mo-data discussed here, a 
reasonable extraction of strength information should not ignore the fluctuating 
quasi continuous part of the cross section. \\
Thus $f_1(E)$ was determined for the 
three Mo-isotopes by using all scattering strength with the exception of the 
nonresonantly scattered photons, whose contribution to the spectra was 
calculated and subtracted. The $f_1$-data for the three Mo isotopes show a clear 
maximum at $\sim$9\,MeV indicating the presence of a pygmy resonance, as it was 
observed at this energy also in $^{90}$Zr and slightly lower in energy in Sn 
isotopes. The 9\,MeV-structure is below $S_n$ in $^{92}$Mo and clearly above 
$S_n$ in $^{100}$Mo; in $^{98}$Mo a cross section irregularity shows up at the 
neutron threshold. Intermediate structures are weakly showing up also between 6 
and 7 MeV (as in $^{118}$Sn and $^{124}$Sn) and eventually also at 11\,MeV (as 
in $^{90}$Zr). This may be considered an indication for a sequence of pygmy 
resonances - not just one.   

\section{Acknowledgements}
Dr.\,P.\,Michel and the ELBE-Crew made these experiments possible with their 
strong commitment to deliver optimum beams. A. Hartmann and W. Schulze provided 
very valuable support during the difficult experiments. Intensive discussions 
with Dr.\,F.\,Becvar, Dr.\,F.\,D\"onau and Dr.\,R.  W\"unsch are gratefully 
acknowledged. The DFG has supported one of us (G.\,R.) under Do466/1-2 during 
the course of the studies presented here.\\

\printindex
\end{document}